\documentclass[letter,onecolumn]{jpsj2}
\usepackage{color}
\usepackage{ulem}
\begin{document}
\title{Superconductivity in Ca$_{1-x}$La$_{x}$FeAs$_2$: A Novel 112-Type Iron Pnictide with Arsenic Zigzag Bonds}

\author{
Naoyuki \textsc{Katayama}$^{1}$\thanks{E-mail address: katayama@mcr.nuap.nagoya-u.ac.jp},
Kazutaka \textsc{Kudo}$^{2}$\thanks{E-mail address: kudo@science.okayama-u.ac.jp},
Seiichiro \textsc{Onari}$^{1}$,
Tasuku \textsc{Mizukami}$^{2}$,
Kento \textsc{Sugawara}$^{1}$,
Yuki \textsc{Sugiyama}$^{1}$,
Yutaka \textsc{Kitahama}$^{2}$,
Keita \textsc{Iba}$^{2}$,
Kazunori \textsc{Fujimura}$^{2}$,
Naoki \textsc{Nishimoto}$^{2}$,
Minoru \textsc{Nohara}$^{2}$, and Hiroshi \textsc{Sawa}$^{1}$
}

\inst{
$^{1}$Department of Applied Physics, Nagoya University, Nagoya 464-8603, Japan\\
$^{2}$Department of Physics, Okayama University, Okayama 700-8530, Japan
}

\abst{
We report superconductivity in the novel 112-type iron-based compound Ca$_{1-x}$La$_{x}$FeAs$_2$.
Single-crystal X-ray diffraction analysis revealed that the compound crystallizes in a monoclinic structure (space group $P2_1$), in which Fe$_2$As$_2$ layers alternate with Ca$_2$As$_2$ spacer layers such that monovalent arsenic forms zigzag chains.
Superconductivity with a transition temperature ($T_{\rm c}$) of 34 K was observed for the $x$ = 0.1 sample,
while the $x$ = 0.21 sample exhibited trace superconductivity at 45 K.
First-principles band calculations demonstrated the presence of almost cylindrical Fermi surfaces,
favorable for the high $T_{\rm c}$ in La-doped CaFeAs$_2$.
}

\kword{Iron-based superconductors, Ca-La-Fe-As, 112-type, CaFeAs$_2$}

\maketitle

Since the discovery of superconductivity with a transition temperature ($T_{\rm c}$) of 26 K in LaFeAsO$_{1-x}$F$_x$,\cite{rf:1} there has been tremendous effort towards synthesizing novel iron pnictide superconductors.\cite{rf:7,rf:8,rf:9,rf:10,rf:11,rf:12,rf:13,rf:14,rf:15,rf:16,rf:17,rf:18,rf:19}
All of the iron pnictide superconductors identified so far consist of a common structural motif, i.e., Fe$_2$As$_2$ layers that are alternately stacked with various kinds of spacer layers. 
Therefore, the central goal for realizing a higher $T_{\rm c}$ has been finding a novel spacer layer that can suitably tune the electronic states of Fe$_2$As$_2$ layers.

Recently, superconductivity has been discovered in Ca$_{10}$(Pt$_{n}$As$_8$)(Fe$_{2-x}$Pt$_x$As$_2$)$_5$, which consists of As-As dimers with a formal electron count of As$^{2-}$ in the spacer layer.\cite{rf:2, rf:3, rf:4, rf:24}
Because of the 4$p^3$ electron configuration of elemental arsenic, arsenic can form various bonding structures:
(i) Isolated arsenic with a formal electron count of As$^{3-}$. 
Examples include $A_3$As ($A$ = Li, Na, and K) and iron-based superconductors.
(ii) Dimerized As-As with a single bond. 
Its formal electron count is As$^{2-}$. 
Sr$_2$As$_2$ and Ca$_{10}$(Pt$_{n}$As$_8$)(Fe$_{2-x}$Pt$_x$As$_2$)$_5$ with As-As dimer bonds in the spacer layer can be categorized here.
(iii) A one-dimensional chain connected by arsenic single bonds with a formal electron count of As$^{-}$.
This category includes KAs as an example. Realizing novel iron-based superconductors with spacer layers composed of complex bonding networks of arsenic such as (iii) has been a longstanding challenge: Shim $et~al.$ have theoretically proposed the hypothetical compound BaFeAs$_2$ (112-type) with spacer layers of the arsenic square network, and suggested that such compounds can be used to examine the role of charge and polarization fluctuations as well as the importance of two-dimensionality in the mechanism of superconductivity.\cite{rf:26} Although the 112-type iron pnictides $AE$FeAs$_2$ ($AE$ = Ca, Sr, Ba) have not yet been synthesized, the isostructural compounds $RET$As$_2$ ($RE$ = rare-earth elements; $T$ = Cu, Ag, Au) have been studied intensively.\cite{rf:27,rf:23}

In this letter, we present a report on the novel 112-type iron-based superconductor Ca$_{1-x}$La$_{x}$FeAs$_2$. 
Although pure CaFeAs$_2$ was not obtained, we found that the substitution of a small amount of La for Ca stabilizes the 112 phase. Thus, Ca$_{1-x}$La$_{x}$FeAs$_2$ was synthesized for the first time. Single-crystal X-ray diffraction analysis revealed that the compound consists of arsenic zigzag bond layers that are composed of arsenic single bonds with a formal electron count of As$^-$. 
This is the first example of an iron-based superconductor belonging to the monoclinic space group $P2_1$ (No. 4). 
Magnetization measurements demonstrate the emergence of bulk superconductivity at 34 K in Ca$_{1-x}$La$_{x}$FeAs$_2$ with a nominal composition of $x$ = 0.10. 
Moreover, resistivity measurements revealed trace superconductivity at 45 K for a nominal composition of $x$ = 0.20, suggesting the possible increase in $T_{\rm c}$ in this compound. 
Furthermore, first-principles band calculations {demonstrated} that the overall appearance of the Fermi surface is similar to those of 1111-type compounds,\cite{rf:25,rf:21} which exhibit high-$T_{\rm c}$ superconductivity with the suitable tuning of electronic states using chemical methods.\cite{rf:19}

\begin{figure*}[t]
\begin{center}
\includegraphics[width=12.3cm]{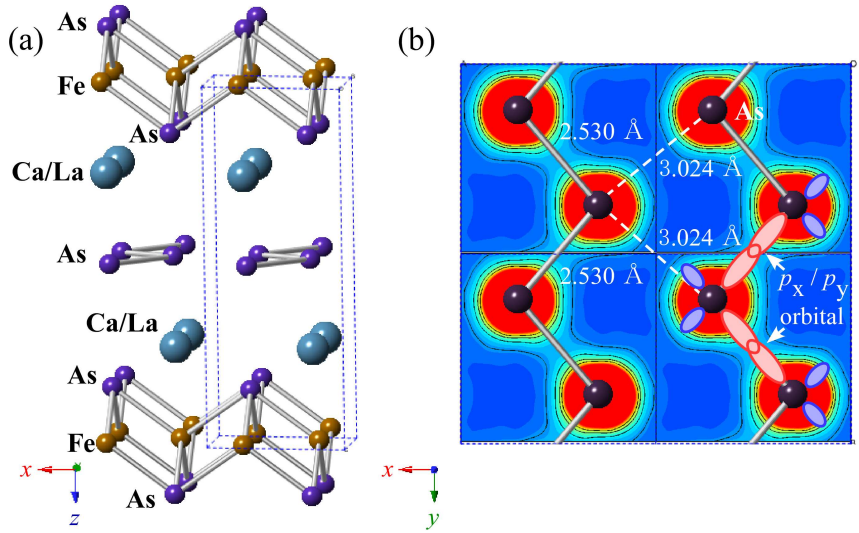}
\caption{\label{fig:Fig1}
(Color online) (a) Crystal structure of Ca$_{1-x}$La$_{x}$FeAs$_2$ with a monoclinic structure having the space group $P2_1$ ($\sharp$4). Blue dotted lines indicate the unit cell. (b) Top view of the arsenic zigzag chains. The unit cell contains two arsenic ions comprising the zigzag chains, which are related by a $2_1$ axis. Arsenic 3$p_x$ and 3$p_y$ orbitals are schematically shown. Orbital phases, which are determined using the crystallographic symmetry operations of the space group, are depicted using colors. The background color contour map shows the charge density distributions obtained by synchrotron X-ray diffraction analysis. The contour lines are drawn from 0 to 1.25 $e$ \AA $^{-3}$ at intervals of 0.2 $e$ \AA $^{-3}$.}
\end{center}
\end{figure*}

Single crystals of Ca$_{1-x}$La$_{x}$FeAs$_2$ with nominal compositions of $x$ = 0.10 and 0.21 were grown by heating a mixture of Ca, La, FeAs, and As powders.
A mixture having a ratio of Ca:La:FeAs:As = $1-x$:$x$:1.14:0.99 was placed in an aluminum crucible and sealed in an evacuated quartz tube. 
The preparation was carried out in a glove box filled with argon gas.
The ampules were heated at 700 $^\circ$C for 3 h, slowly heated to 1100 $^\circ$C, and cooled to 1050 $^\circ$C at a rate of 1.25 $^\circ$C/h, followed by furnace cooling.
Single-crystal X-ray diffraction experiments were performed at SPring-8 BL02B1 (Hyogo, Japan).
Single crystals with a typical dimensions of 30 $\times$ 30 $\times$ 10 $\mu$m$^3$ were used for the BL02B1 experiment.
The X-ray wavelength was 0.52 \AA.
The lattice parameters obtained and the refined conditions are summarized in Table \ref{tab:table1}.
In order to obtain a charge density map using the maximum entropy method (MEM), synchrotron X-ray diffraction data with a reliability factor of 2.06\% in the resolution range of $d$ $\geq$ 0.30 \AA~was used.
The chemical compositions were analyzed by energy-dispersive X-ray spectrometry (EDS).
Electrical resistivity $\rho_{\rm ab}$ (parallel to the $ab$-plane) measurements were carried out by a standard DC four-terminal method in a physical property measurement system (Quantum Design PPMS).
The magnetization $M$ was measured using a SQUID magnetometer (Quantum Design MPMS).

\begin{table}[t]
\begin{center}
 \caption{Data collection and refinement statistics for the synchrotron X-ray structure determination of Ca$_{1-x}$La$_{x}$FeAs$_2$.}
 \label{tab:table1}
 \vspace{3mm}
{\tabcolsep=3mm
 \begin{tabular}{ll}
  \hline
  \hline
Ca$_{1-x}$La$_{x}$FeAs$_2$ &  100 K\\
  \hline
  \hline
 Data Collection & \\
 Crystal System & monoclinic \\
 Space Group & $P2_1$\\
 $a$ (\AA) & 3.94710(10) \\
 $b$ (\AA) & 3.87240(10) \\
 $c$ (\AA) & 10.3210(7) \\
 $\alpha$, $\beta$, $\gamma$ ($^{\circ}$) & 90, 91.415(6), 90\\
 $R_{\rm merge}$ & 0.0469\\
 $I$ / $\sigma$ $I$ & $>$2\\
 Completeness (\%) & 0.742\\
 Redundancy & 8.31\\
 &\\
 Refinement &\\
 Resolution (\AA) & 0.30\\
 No. of Reflections & 2573\\
 $R$1 &0.0206\\
 No. of Atoms & 5\\
 \hline
  \hline
 \end{tabular}}
\end{center}
\end{table}
\begin{table}[hbt]
\renewcommand{\arraystretch}{1}
\caption{
Crystallographic parameters of Ca$_{1-x}$La$_{x}$FeAs$_2$ with the space group $P2_1$ at 100 K for $x$ = 0.195. The site occupancy, atomic coordinates, and thermal parameters of Ca(1) and La(1) were refined. A crystal information file (CIF) of the crystal structure of Ca$_{x}$La$_{1-x}$FeAs$_2$ derived by analysis can be obtained free of charge from The Cambridge Crystallographic Data Centre via www.ccdc.cam.ac.uk. CCDC 961741 contains the crystallographic data for this paper.
}
\label{tab:table2}
\begin{tabular}{lccccc}
\hline
\hline
\multicolumn{5}{c}{Ca$_{1-x}$La$_{x}$FeAs$_2$ for $x$ = 0.195}\\
\hline
\multicolumn{5}{c}{Atomic Position}\\
Site~~  & Occupancy~~ & $x/a$ & $y/b$ & $z/c$ \\
\hline
As(1)~~  & 1~~ & 0.70629(3) & 0.25 & 0.504447(12) \\
As(2)~~  & 1~~ & 0.74134(3) & 0.7497(3) & 0.862554(12) \\
Fe(1)~~  & 1~~ & 0.74993(4) & 0.2492(3) & 0.001257(17) \\
Ca(1)~~  & 0.8055(17)~~ & 0.2272(3) & 0.2448(4) & 0.73154(13) \\
La(1)~~  & 0.1946(11)~~ & 0.2399(4) & 0.2563(6) & 0.72863(16) \\
\hline
\hline
\end{tabular}
\end{table}

\begin{figure}[t]
\begin{center}
\includegraphics[width=6cm]{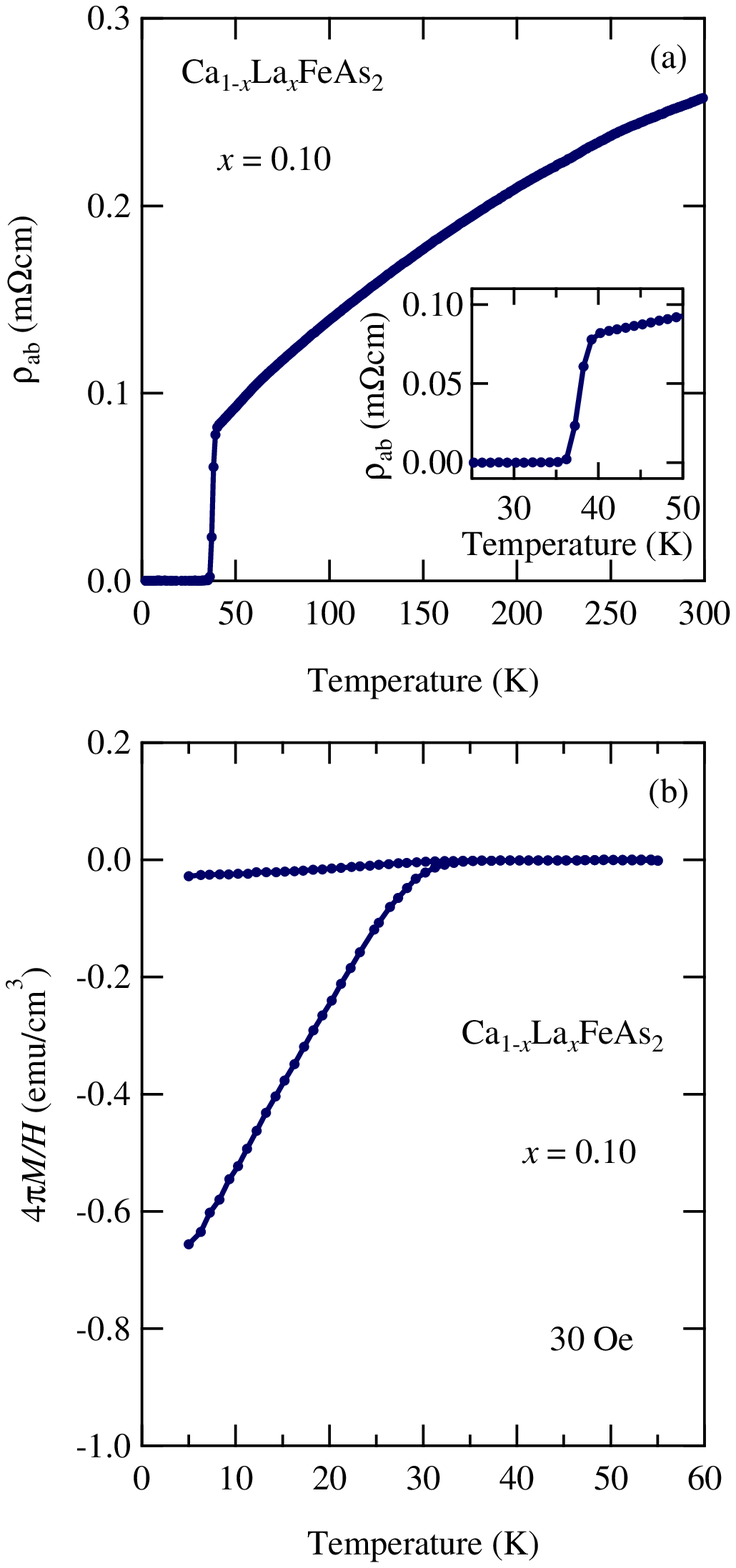}
\caption{\label{fig:Fig2}
(Color online) (a) Temperature dependence of the electrical resistivity parallel to the $ab$-plane, $\rho_{\rm ab}$, for Ca$_{1-x}$La$_{x}$FeAs$_2$ with a nominal composition of $x$ = 0.10. 
The inset shows a magnified view in the vicinity of the superconducting transition. 
(b) Temperature dependence of the magnetization $M$ measured at a magnetic field $H$ of 30 Oe for Ca$_{1-x}$La$_{x}$FeAs$_2$ with a nominal composition of $x$ = 0.10 under zero-field cooling and field cooling.}
\end{center}
\end{figure}

\begin{figure}[t]
\begin{center}
\includegraphics[width=6cm]{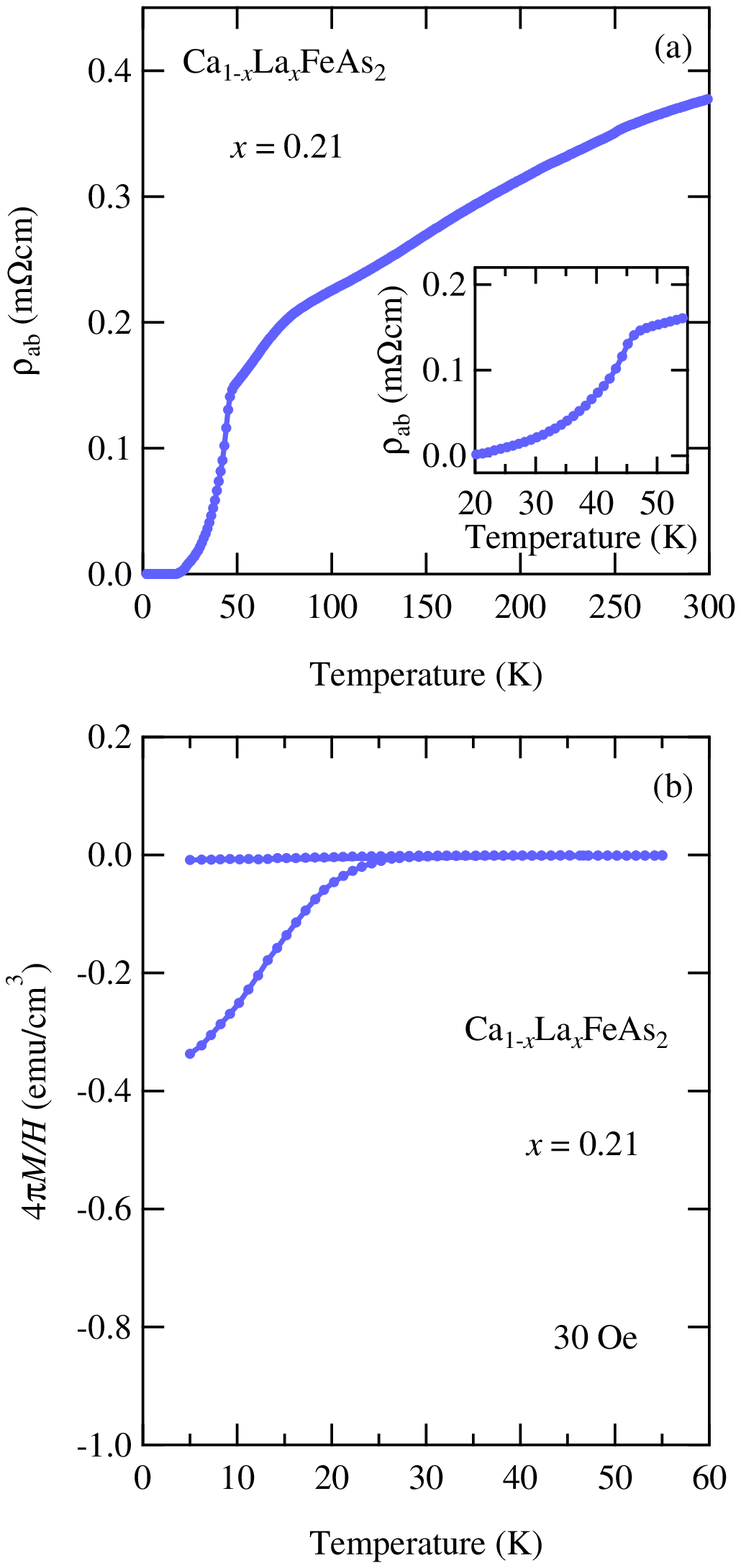}
\caption{\label{fig:Fig3}
(Color online) (a) Temperature dependence of the electrical resistivity parallel to the $ab$-plane, $\rho_{\rm ab}$, for Ca$_{1-x}$La$_{x}$FeAs$_2$ with nominal $x$ = 0.21. The inset shows a magnified view in the vicinity of the superconducting transition. (b) Temperature dependence of the magnetization $M$ measured at a magnetic field $H$ of 30 Oe for Ca$_{1-x}$La$_{x}$FeAs$_2$ with a nominal composition of $x$ = 0.21 under zero-field cooling and field cooling.}
\end{center}
\end{figure}

Single-crystal X-ray diffraction analysis revealed that Ca$_{1-x}$La$_{x}$FeAs$_2$ crystallizes in a monoclinic structure with the space group $P2_1$ (No. 4), in contrast to conventional iron pnictide superconductors belonging to tetragonal or orthorhombic space groups. 
High-temperature X-ray diffraction experiments indicated that the monoclinic structure is stable up to 450 K. 
As shown in Fig. \ref{fig:Fig1}(a), Ca$_{1-x}$La$_{x}$FeAs$_2$ consists of alternately stacked Fe$_2$As$_2$ and arsenic layers. 
The distance between adjacent Fe$_2$As$_2$ layers is $d$ $\simeq$ 10.3 \AA, which is comparable to $d$ $\simeq$ 10 \AA~in Ca$_{10}$(Pt$_{n}$As$_8$)(Fe$_{2-x}$Pt$_x$As$_2$)$_5$\cite{rf:2,rf:4,rf:24,rf:3} and slightly larger than $d$ $\simeq$ 8.7 \AA~in LaFeAsO\cite{rf:1} and $d$ $\simeq$ 9.0 \AA~in SrFeAsF.\cite{rf:22} 
The presence of arsenic layers is the most notable feature in the present structure. As shown in Fig. \ref{fig:Fig1}(b), the in-plane adjacent As-As distances can be classified as short and long. 
The short As-As distance is approximately 2.53 \AA, which is almost equal to the As-As single bond length of 2.52 \AA~in elemental As. Short As-As bonds form one-dimensional zigzag chains along the $b$-axis, as shown in Fig. \ref{fig:Fig1}(b). 
The long As-As distance is approximately 3.02 \AA, which corresponds to the interchain distance.

The presence of chemical bonds between short As-As bonds was confirmed in the MEM charge density map obtained using synchrotron X-ray diffraction analysis.  
The color contour map in Fig. \ref{fig:Fig1}(b) clearly shows the charge density distributions of regular zigzag arsenic chains. In contrast, significant charge distributions are absent between interchain arsenic ions, indicating the absence of chemical bonds between them.

The electrical resistivity and magnetization data revealed that Ca$_{1-x}$La$_{x}$FeAs$_2$ is a novel iron-based superconductor.
As shown in Fig. \ref{fig:Fig2}(a), the electrical resistivity $\rho_{\rm ab}$ of Ca$_{1-x}$La$_{x}$FeAs$_2$ with a nominal composition of $x$ = 0.10 (Ca$_{0.84}$La$_{0.16}$Fe$_{1.03}$As$_{2.37}$, as determined from EDS) exhibits a sharp drop below 39 K, characteristic of a superconducting transition.
Zero resistivity is observed at 36 K, and the 10--90\% transition width is estimated to be 2.4 K.
Bulk superconductivity in this sample was clearly demonstrated by the temperature dependence of the magnetization $M$, as shown in Fig. \ref{fig:Fig2}(b).
$M(T)$ exhibits diamagnetic behavior below 34 K.
The shielding volume fraction corresponds to 66\% for perfect diamagnetism.

Note that a phase with a higher $T_{\rm c}$ is suggested in the present compound.
As shown in Fig. \ref{fig:Fig3}(a), $\rho_{\rm ab}(T)$ for Ca$_{1-x}$La$_{x}$FeAs$_2$ with a nominal composition of $x$ = 0.21 (Ca$_{0.81}$La$_{0.19}$Fe$_{1.03}$As$_{2.34}$, as determined from EDS) exhibits the onset of superconductivity at 45 K, which is much higher than that of the sample with $x$ = 0.10.
The superconducting transition is broad, and zero resistivity is observed at 25 K.
This temperature of 25 K is consistent with $T_{\rm c}$ determined from the magnetization; a visible diamagnetic signal is obtained below 25 K, whereas no signal is observed at approximately 45 K, as shown in Fig. \ref{fig:Fig3}(b).
The trace superconductivity at 45 K suggests that the system still possesses a large potential for the increase in $T_{\rm c}$.

The present compound may be written as the chemical formula (Ca$^{2+}_{1-x}$La$^{3+}_x$)(Fe$^{2+}_2$As$^{3-}_2$)$_{1/2}$As$^{-}$$\cdot$$xe^-$ with an excess charge of $xe^{-}$ injected into the Fe$^{2+}_2$As$^{3-}_2$ layers. Here, the arsenic zigzag bond layers are composed of As$^-$ ions with 4$p^4$ electronic states. 
Owing to the anisotropic crystal field around the arsenic ions, the threefold degeneracy of the 4$p$ orbitals of the As$^{-}$ ions should be inherently lifted. 
Although a large van der Waals gap appears along the out-of-plane directions, the arsenic ions are closely packed along the in-plane directions, resulting in a stabilized $p_z$ orbital with two electrons and destabilized $p_x$ and $p_y$ orbitals with one unpaired electron. 
Because the $p_x$ and $p_y$ orbitals spread in the in-plane directions with rectangular coordinates, we expect that the zigzag bonds are formed by the chemical bond between adjacent arsenic atoms oriented by the $p_x$ and $p_y$ orbitals, as shown in Fig. \ref{fig:Fig1}(b). 
The chemical bonds composed of zigzag chains were experimentally observed in the charge density map obtained using MEM, as presented above.

Such complex arsenic network structures also appear in other 112-type layered arsenides with the chemical formula $LnT$As$_2$ ($Ln$ = lanthanide, $T$ = Ag, Au).\cite{rf:23} 
Although the stacking structure and orthorhombic space group of $Pmcn$ (No. 62) are in sharp contrast to those of the present Ca$_{1-x}$La$_{x}$FeAs$_2$, SmAuAs$_2$ has a similar structural motif of Au$_2$As$_2$ layers and arsenic zigzag bond layers. 
More remarkably, other types of arsenic networks can also be realized. LaAgAs$_2$, which crystallizes in the orthorhombic space group $Pmca$ (No. 57), consists of Ag$_2$As$_2$ layers and arsenic cis-trans chain layers. 
The arsenic cis-trans chain layers have not yet been realized in iron-based superconductors to date, allowing us to take advantage of the flexible network structures of arsenic according to the 4$p^3$ electronic states which may provide further opportunities to discover novel iron-based superconductors.

\begin{figure}[hbt]
\begin{center}
\includegraphics[width=8cm]{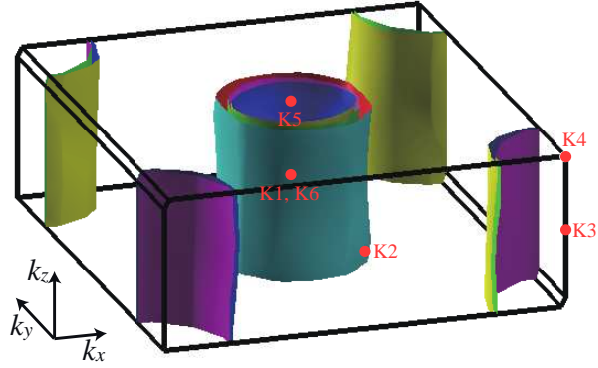}
\caption{\label{fig:Fig4}
(Color online) Fermi surface of CaFeAs$_2$ calculated using the WIEN2k package. Here, the reciprocal coordinates for K1--K6 are, K1(0 0 0), K2(0 -0.5 0), K3(0.5 -0.5 0), K4(0.49558 -0.5 0.46800), K5(0 0 0.5), and K6(0 0 0).}
\end{center}
\end{figure}

Finally, we present first-principles calculations of the present Ca$_{1-x}$La$_{x}$FeAs$_2$ using the WIEN2k package.\cite{rf:20} 
For simplification, we performed the calculation for CaFeAs$_2$ without replacing Ca with La, although a La-free sample has not been experimentally synthesized. 
The structural parameters in Table \ref{tab:table2} were employed for the calculation. 
Looking at the density of states, we observe a strong Fe-3$d$ character located near the Fermi level. While conventional iron-based superconductors crystallize in tetragonal or orthorhombic structures, the present Ca$_{1-x}$La$_{x}$FeAs$_2$ belongs to the monoclinic space group $P2_1$. 
However, the $\beta$ angle is close to 90 deg, resulting in an almost cylindrical Fermi surface and the absence of significant modifications by symmetry lowering, as shown in Fig. \ref{fig:Fig4}. 
The overall appearance is similar to that of LaFeAsO\cite{rf:25} and SrFeAsF,\cite{rf:21} which exhibit high-$T_c$ superconductivity with a transition temperature of $\sim$ 55 K with the suitable tuning of electronic states using chemical methods.\cite{rf:19} 
This similarity seems to be strongly related to the emergence of high-$T_{\rm c}$ superconductivity in the present material. 
Moreover, our preliminary band calculation results also show that the modification of the Fermi surface is small when substituting La for Ca, indicating that the development of chemical methods of optimizing the carrier concentrations will increase the superconducting transition temperature further.

In summary, we have discovered a novel 112-type iron pnictide superconductor, Ca$_{1-x}$La$_{x}$FeAs$_2$, with arsenic zigzag bond layers.
The sample with a nominal composition of $x$ = 0.10 exhibited bulk superconductivity at $T_{\rm c}$ = 34 K.
Moreover, the sample with a nominal composition of $x$ = 0.21 exhibited the onset of trace superconductivity at $T_{\rm c}$ = 45 K.
First-principles calculations clearly indicated a cylindrical-like Fermi surface, which is similar to those of 1111-system superconductors such as LaFeAsO and SrFeAsF. We expect that the optimization of the carrier concentration using chemical procedures will further increase $T_c$.

\begin{acknowledgements}
We thank W. Yamada for his technical assistance. The work at Nagoya University was supported by a Grant-in-Aid for Scientific Research (No. 23244074). The work at Okayama University was partially supported by a Grant-in-Aid for Scientific Research (C) (No. 25400372) from Japan Society for the Promotion of Science (JSPS) and the Funding Program for World-Leading Innovative R\&D on Science and Technology (FIRST Program) from JSPS.
Some of the magnetization measurements were carried out at the Advanced Science Research Center, Okayama University. The synchrotron radiation experiments performed at BL02B1 in SPring-8 were supported by the Japan Synchrotron Radiation Research Institute (JASRI) (Proposal Nos. 2012A0083, 2012B0083 and 2013A0083).
\end{acknowledgements}

$Note~added$ $-$ Yakita $et~al.$ have recently reported superconductivity in similar compounds with arsenic layers.\cite{rf:28}


\begin{thebibliography}{99}
\bibitem{rf:1} Y. Kamihara, T. Watanabe, M. Hirano, and H. Hosono: J. Am. Chem. Soc. \textbf{130} (2008) 3296.
\bibitem{rf:7} J. H. Tapp, Z. Tang, B. Lv, K. Sasmal, B. Lorenz, P. C. W. Chu, and A. M. Guloy: Phys. Rev. B \textbf{78} (2008) 060505(R).
\bibitem{rf:8} M. Rotter, M. Tegel, and D. Johrendt: Phys. Rev. Lett. \textbf{101} (2008) 107006.
\bibitem{rf:9} S. Matsuishi, Y. Inoue, T. Nomura, H. Yanagi, M. Hirano, and H. Hosono: J. Am. Chem. Soc. \textbf{130} (2008) 14428.
\bibitem{rf:10} K. Kudo, K. Iba, M. Takasuga, Y. Kitahama, J. Matsumura, M. Danura, Y. Nogami, and M. Nohara: Sci. Rep. \textbf{3} (2013) 1478.
\bibitem{rf:11} N. Kawaguchi, H. Ogino, Y. Shimizu, K. Kishio, and J. Shimoyama: Appl. Phys. Express \textbf{3} (2010) 063102.
\bibitem{rf:12} X. Zhu, F. Han, G. Mu, P. Cheng, B. Shen, B. Zeng, and H.-H. Wen: Phys. Rev. B \textbf{79} (2009) 220512(R).
\bibitem{rf:13} H. Ogino, K. Machida, A. Yamamoto, K. Kishio, J. Shimoyama, T. Tohei, and Y. Ikuhara: Supercond. Sci. Technol. \textbf{23} (2010) 115005.
\bibitem{rf:14} P. M. Shirage, K. Kihou, C.-H. Lee, H. Kito, H. Eisaki, and A. Iyo: Appl. Phys. Lett. \textbf{97} (2010) 172506.
\bibitem{rf:15} H. Ogino, Y. Katsura, S. Horii, K. Kishio, and J. Shimoyama: Supercond. Sci. Technol. \textbf{22} (2009) 085001.
\bibitem{rf:16} Y. Matsumura, H. Ogino, S. Horii, Y. Katsura, K. Kishio, and J. Shimoyama: Appl. Phys. Express \textbf{2} (2009) 063007.
\bibitem{rf:17} S. Sato, H. Ogino, N. Kawaguchi, Y. Katsura, K. Kishio, J. Shimoyama, H. Kotegawa, and H. Tou: Supercond. Sci. Technol. \textbf{23} (2010) 045001.
\bibitem{rf:18} H. Ogino, S. Sato, K. Kishio, J. Shimoyama, T. Tohei, and Y. Ikuhara: Appl. Phys. Lett. \textbf{97} (2010) 072506.
\bibitem{rf:19} Z. A. Ren, W. Lu, J. Yang, W. Yi, X. L. Shen, Z. C. Li, G. C. Che, X. L. Dong, L. L. Sun, F. Zhou, and Z. X. Zhao: Chin. Phys. Lett. \textbf{25} (2008) 2215.
\bibitem{rf:2} S. Kakiya, K. Kudo, Y. Nishikubo, K. Oku, E. Nishibori, H. Sawa, T. Yamamoto, T. Nozaka, and M. Nohara: J. Phys. Soc. Jpn. \textbf{80} (2011) 093704.
\bibitem{rf:4} N. Ni, J. M. Allred, B. C. Chan, and R. J. Cava: Proc. Natl. Acad. Sci. \textbf{108} (2011) E1019.
\bibitem{rf:24} C. L\"{o}hnert, T. St\"{u}rzer, M. Tegel, R. Frankovsky, G. Friederichs, and D. Johrendt: Angew. Chem. Int. Ed. \textbf{50} (2011) 9195. 
\bibitem{rf:3} M. Nohara, S. Kakiya, K. Kudo, Y. Oshiro, S. Araki, T. C. Kobayashi, K. Oku, E. Nishibori, and H. Sawa: Solid State Commun. \textbf{152} (2012) 635.
\bibitem{rf:26} J. H. Shim, K. Haule, and G. Kotliar: Phys. Rev. B \textbf{79} (2009) 060501(R). 
\bibitem{rf:27} M. Brylak, M. H. M\"{o}ller, and W. Jeitschko: J. Solid State Chem. \textbf{115} (1995) 305. 
\bibitem{rf:23} D. Rutzinger, C. Bartsch, M. Doerr, H. Rosner, V. Neu, Th. Doert, and M. Ruck: J. Solid State Chem. \textbf{183} (2010) 510.
\bibitem{rf:25} D. J. Singh and M.-H. Du: Phys. Rev. Lett. \textbf{100} (2008) 237003.
\bibitem{rf:21} I. A. Nekrasov, Z. V. Pchelkina, and M. V. Sadovskii: JETP Lett. \textbf{88} (2008) 679. 
%
\bibitem{rf:22} G. Wu, Y. L. Xie, H. Chen, M. Zhong, R. H. Liu, B. C. Shi, Q. J. Li, X. F. Wang, T. Wu, Y. J. Yan, J. J. Ying, and X. H. Chen: J. Phys.: Condens. Matter \textbf{21} (2009) 142203. 
\bibitem{rf:20} P. Blaha, K. Schwarz, G. K. H. Madsen, D. Kvasnicka, and J. Luitz: $Wien2k:$ $An$ $Augmented$ $Plane$ $Wave$ $+$ $Local$ $Orbitals$ $Program$ $for$ $Calculating$ $Crystal$ $Properties$, Vienna University of Technology, Wien, Austria, 2001.
\bibitem{rf:28} H. Yakita, H. Ogino, T. Okada, S. J. Singh, A. Sala, A. Yamamoto, K. Kishio, T. Tohei, Y. Ikuhara, H. Fujihisa, Y. Gotoh, H. Eisaki, and J. Shimoyama:  Meet. Abstr. Jpn. Soc. Appl. Phys. \textbf{17p-C8-8}, (2013) 11-072. 
\end{thebibliography}
\end{document}